\documentclass[fleqn,copyright,submission]{eptcs}
\providecommand{\event}{Refine 2011}  
\usepackage{breakurl}                

\usepackage{amssymb}
\usepackage{amsmath}
\usepackage{graphicx}
\usepackage{comment}
\usepackage{url}

\title{Formalising the Continuous/Discrete Modeling Step}

\author{Richard Banach\thanks{The majority of the work reported
in this paper was done while the first author was a visiting researcher
at the Software Engineering Institute at East China Normal University.
The support of ECNU is gratefully acknowledged.}
\institute{School of Computer Science, University of Manchester,\\
Oxford Road, Manchester, M13 9PL, U.K.}
\email{banach@cs.man.ac.uk}
\and Huibiao Zhu\thanks{Huibiao Zhu is supported by the National Basic
Research Program of China (No. 2011CB302904), the National Natural Science
Foundation of China (No. 61061130541), China HGJ Significant Project
(No. 2009ZX01038-001-07), and Doctoral Program Foundation of Institutions
of Higher Education of China (No. 200802690018).}
\qquad\qquad Wen Su
\institute{Software Engineering Institute, East China Normal University,\\
3663 Zhongshan Road North, Shanghai 200062, P.R. China.}
\email{\{hbzhu,wensu\}@sei.ecnu.edu.cn}
\and Runlei Huang
\institute{Alcatel-Lucent Shanghai Bell, 388 Ningqiao Road,\\
Pudong Jinqiao, Shanghai 201206, P.R. China.}
\email{runleihuang@alcatel-sbell.com.cn}
}

\def\titlerunning{Formalising the Continuous/Discrete Modeling Step}
\def\authorrunning{R. Banach, H. Zhu, W. Su, R. Huang}

\begin{document}

\maketitle

\begin{abstract}
Formally capturing the transition from a continuous model to a
discrete model is investigated using model based refinement
techniques. A very simple model for stopping (eg.~of a train)
is developed in both the continuous and discrete domains. The
difference between the two is quantified using generic results
from ODE theory, and these estimates can be compared with the exact
solutions. Such results do not fit well into a conventional model
based refinement framework; however they can be accommodated into
a model based retrenchment. The retrenchment is described, and
the way it can interface to refinement development on both
the continuous and discrete sides is outlined. The approach
is compared to what can be achieved using hybrid systems techniques.
\end{abstract}

\section{Introduction}
\label{intro}

Conventional model based formal refinement technologies (see for example
\cite{roever:98,boiten:01,sekerinski:98,abrial:96,potter:96,woodcock:96,abrial:10})
are based on purely discrete mathematical and logical concepts. These
turn out to be ill suited to modeling and formally developing applications
whose usual models are best expressed using continuous mathematics.
Nevertheless, many such applications, control systems in particular,
are these days implemented using digital techniques. So there is a
mismatch between continuous modeling and discrete development techniques.

In this paper we tackle this mismatch head on. Although traditional
model based refinement is too exacting to straddle the continuous to
discrete demarcation line, a judicious weakening of it, retrenchment,
proves to be adaptable enough to do the job, which we show. Importantly,
retrenchment techniques interface well with refinement, so that a
development starting from continuous and ending at discrete can be
captured in an integrated way.

In this paper we tackle the continuous to discrete issue by taking
a simple running example, one that can be solved fully by analytic
means in both the continuous and discrete domains, and tracing it
through the critical formal development step. We start with a continuous
control problem: bringing an object (eg.~a train) to a halt. This is
formulated as a continuous control problem, and given the (deliberately
chosen) simplicity of the problem, an exact solution is presented. In
reality, continuous control is implemented these days via digital
controllers. These periodically read inputs and recompute outputs
at multiples of a sampling interval during the dynamics. In this sense,
the control becomes discretized, although the discretized control
is obviously still played out in the continuous real world. We thus
remodel the continuous problem as a discrete control problem, and
derive a formal description of the discretization step via a suitable
retrenchment, drawing on rigorous results from the theory of ordinary
differential equations (ODEs) to supply the justification. Given the
limited size of this paper, our technical focus is on this critical
step, and the remainder of the development (comprising the associated
refinements either side of it) is sketched rather than treated in detail.
The latter is a task for which a fuller treatment will be given in the
extended version of the paper.

The rest of the paper is as follows. We start in Section \ref{sec-RelWork}
by describing relevant existing work in the hybrid systems domain and how
it contrasts with our own approach, after which we get down to details.
Section \ref{sec-TrStReq} then formulates our train stopping problem as a
conventional open loop continuous control problem. Section \ref{sec-CtDisc}
then describes the discretization of the control problem using a simple
zero order hold strategy. In Section \ref{sec-ASMRefRet} we review what
we need of ASM refinement and retrenchment in a form suitable for our
problem. Section \ref{sec-FormCtDisc} then shows how our earlier
discretization process can be captured using a suitable retrenchment,
citing the needed ODE results. Section \ref{sec-ContDes} sketches how
all this can fit into a wider formal development strategy, in which the
greater flexibility of retrenchment can be combined with the stronger
guarantees offered by refinement via the Tower Pattern \cite{BaJe:09,Jes:05}.
Section \ref{sec-concl} concludes.

\section{Related Work}
\label{sec-RelWork}

The relationship between continuous and discrete transition systems
has long been a topic for investigation in the hybrid systems field.
Earlier work includes \cite{AlCouHenHo:93,Henz:06,AlurDill:94,He:94};
also, the International Conference on Hybrid Systems: Computation and
Control, has been the venue for a large amount of research in this area.
A more recent reference is \cite{Tab:09}.

Hybrid systems are dynamical systems that mix smooth, continuous
transitions with discrete, discontinuous ones. The major focus in
this field has been the automatic verification of properties of such
systems. Obviously, such verification demands the representation of
the systems in question in discrete and finite terms, whether by means
of an explicitly constructed finite state space (which is manipulated
directly), or a state space whose states arise via the symbolic
representation of the less tractable state space of a previously
constructed underlying system (which is manipulated symbolically).

The main tool for bringing an intractable state space within the
scope of computable techniques is the equivalence relation. Regions
of the state space are gathered into equivalence classes, and a
representation of these equivalence classes (whether as individual
elements in a simple approach, or as symbolic expressions that denote
the equivalence class in question) constitutes the state space of the
abstraction. Transitions between these states are introduced to mirror
the behaviour of the underlying system. The properties of interest can
then be checked against the abstract system. For instance, properties
that can be expressed as reachability properties fall within the scope
of model checking approaches that are applied to the abstraction.

Of course what has been constructed thereby is a (bi)simulation, and a
major strand of hybrid systems research is the investigation of such
(bi)simulations. The same remarks apply when there is an external control
applied to the systems.

One disadvantage of the above approach is the frequent reliance on
brittle properties of the studied systems. Put most simply, a number
of techniques rely on the parameters of the problem falling within a
subset of measure zero of the parameter space. Real systems can never
hit such small targets reliably. Equally, the simulation relations
studied can also be just as brittle. To alleviate this, and to address
other issues of interest, the notion of \textit{approximate (bi)simulation}
has been studied in recent years (\cite{Tab:09} gives a good
introduction). Here, instead of defining the simulation relation $R(u,v)$
between an abstract state $u$ and a concrete state $v$ as a simple predicate
on states, it is defined via a distance function $\mathbf{d}$ as
$R_{\epsilon}(u,v) \equiv \mathbf{d}(f(u),v) \leq \epsilon$, where
$f$ is a precise relationship between the two state spaces which is
in some sense ``semantically natural'' (we don't have space to elaborate
on this aspect here). For bisimulation you need a symmetrical arrangement
of course.

(Bi)simulation depends on assuming the appropriate relation between the
two before-states and re-establishing it in the after-states of suitable
pairs of transitions. To preserve a relationship based on distance, the
dynamics needs to be inherently \textit{stable}. The obvious centre of
attention thus becomes stable control systems, normally \textit{linear}
stable control systems, because of their calculational tractability.
These are discussed in very many places,
eg.~\cite{Ogata:08,DorBish:10,DutThBa:97,DadHou:95,ahmed:06,sontag:98,barnett:75,Ants:06}.

In a stable system all trajectories converge to a single point, so the
distance between two trajectories decreases monotonically; hence a
simulation relation based on distance between trajectories is maintained.
But although most systems are designed to be stable in this sense, some are
not, and there can be parts of a system phase space in which trajectories
diverge rather than converge, without rendering the system useless. Below,
we treat in detail a very simple example which happens to be unstable in
the sense just discussed. We know it is not stable because we solve it
exactly.

Also, in the usual hybrid systems literature, it is normal that the discrete
approximation to a given system is manufatured from it (eg.~by constructing
equivalence classes, as indicated above). In our approach, by contrast, we
take a more ``off the shelf'' attitude to discretization, analysing a
straightforward ``zero order hold'' version of the continuous system
(in which the new output values to be sent to the actuators are recalculated
at regular intervals, and the new values are ``held'' for the duration of
the next inteval\footnote{``Zero order'' refers here to ``holding'' the
output value constant throughout the interval, in contrast to a higher order
hold which would use a suitably designed higher order polynomial.})
rather than something extracted from an analysis of the original system.
In this sense our approach is closer to conventional engineering practice,
since it is directed at the typical practical approach. Of course these two
ways of doing things are not mutually exclusive: the parameters of the
zero order hold may fall within the parameters of a discrete approximation
extracted by analysis of the original system, and \textit{vice versa}.
Finally, our approach is via retrenchment, one consequence of which is
that our analysis is not confined to the purely stable case. In effect,
the greater expressiveness of retrenchment permits (the analogue of)
the simulation relation mentioned above, to increase its permitted margin
of error, as well as to decrease it, although this emerges indirectly.

\section{Train Stopping: a Continuous Control System}
\label{sec-TrStReq}

Our target application domain is control problems in the railway sphere.
In this paper we have train stopping as a specific case study. Of course,
in reality, train position control is a complex problem \cite{Su:11,CBTC},
relying on the co-operation of many mechanisms to achieve a reliable outcome,
and we do not have the space to deal with all these aspects and their subtle
interactions. Instead we focus on a single technical issue ---the relationship
between a continuous control problem and its discrete counterpart--- in a
very simple way, commenting on the extreme simplicity below.

Suppose a train, of mass $M$, is traveling at its cruise velocity $V$,
when it needs to stop. We assume that a linearly increasing deceleration
rate $a$ is appropriate. (It has to be said here that our notion of
appropriateness is not quite the usual one. Rather than usability or
any similar consideration governing our choice, simplicity is the priority.
A constant deceleration would have been even simpler --- unfortunately the
zero order hold approximation to constant deceleration is identical to it,
trivialising our problem.) To bring the train to a standstill using linearly
increasing deceleration, a force $F=-Mat$ (where $t$ is time) has to be applied,
by Newton's Law. We will assume that $M$ is known, so that we can focus on just
the kinematic aspects.

A cursory knowledge of kinematics is enough to reveal that
under linear deceleration, the deceleration, distance and stopping time
are linked. We suppose that
there is a single stopping episode, which starts at time $0$ and at $x$
position $0$, and which ends at time $T_{Stop}$, with the train having
traveled to position $x=D$. Representing time derivatives with a dot, if
$v$ is the velocity, then we know that
\begin{align}
\dot{v} &~=~ -a t & v(0) &~=~ V & v(T_{Stop}) &~=~ 0   \label{eq-v-dot=-a}
\end{align}
Regarding the distance traveled $x$, we know that
\begin{align}
\dot{x} &~=~ v & x(0) &~=~ 0 & x(T_{Stop}) &~=~ D       \label{eq-x-dot=v}
\end{align}
Integrating these, rapidly brings us to
\begin{align}
V &~=~ \frac{1}{2} a T_{Stop}^2 &
D &~=~ V T_{Stop} - \frac{1}{3!} a T_{Stop}^3 ~=~ \frac{2}{3} V T_{Stop}
                                                           \label{eq-V=aT}
\end{align}

We now recast the above as a control theory problem. At the introductory
level, control theory is usually developed in the frequency domain
\cite{Ogata:08,DorBish:10,DutThBa:97,DadHou:95}, because of the
relative simplicity and perspicuity of the design techniques in
that domain. However, for results sufficiently rigorous to interface
to formal techniques, we need to go to the state space formulation
favoured by more mathematically precise treatments
\cite{ahmed:06,sontag:98,clarke:97,clarke:87,barnett:75,Ants:06}.
In the state space picture, the system consists
of a number of state variables, and their evolution is governed by
a corresponding number of first order differential equations. State
variables and differential equations mirror the states and transition
systems of model based refinement formalisms sufficiently closely that
we can hope to make a connection between them.

To use the first order framework in our example, the state has to
consist of both the position $x(t)$ and the velocity $v(t)$. So we
get the state vector
\begin{align}
\pmb{x}(t) ~=~ \begin{bmatrix} x(t) \\ v(t) \end{bmatrix}
                                                      \label{eq-state-def}
\end{align}
The dynamics of the system is captured in the equation\footnote{It is
clear that when (\ref{eq-state-dot}) is expressed as a linear control law
(with external control signal), the linear part has only zero eigenvalues.
Thus it is not stable in the usual (Liapunov) sense.}
\begin{align}
\dot{\pmb{x}}(t) ~=~ \begin{bmatrix} \dot{x}(t) \\ \dot{v}(t) \end{bmatrix}
                ~=~ \pmb{f}(\dot{x}(t),u(t))
                ~=~ \begin{bmatrix} v(t) \\ u(t) \end{bmatrix}
                                                      \label{eq-state-dot}
\end{align}
where
\begin{align}
u(t) ~=~ -a t                                             \label{eq-cont-u}
\end{align}
is the external control control signal.
We also have the initial condition
\begin{align}
\pmb{x}(0) ~=~ \begin{bmatrix} 0 \\ V \end{bmatrix}  \label{eq-state-init}
\end{align}

\section{From Continuous Control to Discrete Control}
\label{sec-CtDisc}

To truly implement a continuous control model, such as our case study,
requires analogue apparatus. In the highly digitized world of today,
hardly any such systems are built. Instead, continuous control designs
are discretized, and it is the corresponding digital control systems
that are implemented.

The digital approach to control has many parallels with the continuous
case --- in the frequency domain the main difference is
the use of the $z$-transform rather than the Laplace transform. The
state based picture too boasts many parallels, with first order
difference equations replacing first order differential equations
\cite{FaVi:09,FrPoWo:98,Para:96,Kuo:92}.

In this section we examine a discrete counterpart of the previous
continuous control problem, in preparation for a formal reappraisal
in the next section. One advantage of the extreme simplicity of our example,
is that it admits an analytic solution in both continuous and discrete
domains, enabling an incisive evaluation to be made later, of the reappraisal
in Section \ref{ssec-Corrob}.

The starting point for our problem remains as before: the train, traveling
at velocity $V$, needs to stop after time $T_{Stop}$, having
gone a distance $D_D$.\footnote{We will use a subscript `$D$' to indicate
quantities in the discretized model that differ from their continuous
counterparts.} Instead of doing so continuously though, it will do it in
a number of discrete episodes. For this purpose, let us assume that $T_{Stop}$
is divided into $N$ short periods, each of length $T$, so that
\begin{align}
T_{Stop} ~=~ N T                                     \label{eq-TStop=NT}
\end{align}
Our discretization scheme will be based on a zero order hold, in which
the same control input value is maintained throughout an individual time
period. The counterpart of the linear deceleration rate $a$ of the continuous
treatment, will be a piecewise constant deceleration, with the constant
rate decreasing by an additional multiple of a constant $a_D$ after each
time interval of length $T$.

Calling the discretized velocity variable $v_D$, we have for the
acceleration
\begin{align}
\dot{v}_D(t) ~=~ - k a_D T                              \label{eq-dot-vD}
\end{align}
where
\begin{align}
k ~=~ \left\lceil \frac{t}{T} \right\rceil                   \label{eq-k}
\end{align}
and $k$ ranges over the values $1 \ldots N$. If we set, for a general $t$,
\begin{align}
\delta t_k ~=~ t - (k-1) T ~=~ t - \left\lfloor \frac{t}{T} \right\rfloor T
                                                     \label{eq-delta-t}
\end{align}
then recalling that the initial velocity is $V$, provided
$(k-1) T < t < k T$, the velocity during the $k$'th period is
\begin{align}
v_D(t) ~=~ V - a_D T^2 - 2 a_D T^2 - \ldots -
           (k-1) a_D T^2 - k a_D T \delta t_k          \label{eq-vD(k)}
\end{align}
Since the final velocity is zero, we derive
\begin{align}
V ~=~ a_D T^2 + 2 a_D T^2 + \ldots + N a_D T^2
  ~=~ \frac{1}{2} a_D T^2 N(N+1)                         \label{eq-VD}
\end{align}
Knowing the velocity, we can integrate again, and work out the distance
traveled. Calling the displacement in the discretized world $x_D$, the
contribution to $x_D$ during the period $(k-1) T < t < k T$ comes out as
\begin{align}
   (V - a_D T^2 - 2 a_D T^2 - \ldots - (k-1) a_D T^2) \delta t_k
           - \frac{1}{2} k a_D T \delta t_k^2          \label{eq-xD(k)}
\end{align}
Thus for the total distance we find
\begin{align}
D_D &~=~ N V T
         - a_D T^3 \sum^{N-1}_{k=1} (N-k) k
         - \frac{1}{2} a_D T^3 \sum^N_{k=1} k                  \notag \\
    &~=~ V T_{Stop} - \frac{1}{12} a_D T^3 (2 N^3 + 3 N^2 + N)
                                                     \label{eq-DD-final}
\end{align}
Both (\ref{eq-VD}) and (\ref{eq-DD-final}) feature $a_D$. Substituting
the $a_D$ value from (\ref{eq-VD}) into (\ref{eq-DD-final}) gives
\begin{align}
D_D &~=~ V T_{Stop} \left[ 1 - \frac{2 N^2 + 3 N + 1}{6 N^2 + 6 N} \right]
      ~=~ \frac{2}{3} V T_{Stop} \left[ 1 - \frac{1}{4N} + O(N^{-2}) \right]
                     \label{eq-DVT-no-a}
\end{align}
We see that (\ref{eq-DVT-no-a}) for $D_D$ contains an $O(1/N)$ correction
compared with (\ref{eq-V=aT}) for $D$ (assuming we keep $V$ and
$T_{Stop}$ the same). This is because we have an extra constraint
generated by the requirement that $T_{Stop}$ is an integral multiple of $T$,
making the problem overconstrained if we wished $D$ and $D_D$ to be the
same.

Recasting the preceding as an initial value first order system along the
lines of (\ref{eq-state-def})-(\ref{eq-state-init}) is not hard. The
state vector is
\begin{align}
\pmb{x}_D(t) ~=~ \begin{bmatrix} x_D(t) \\ v_D(t) \end{bmatrix}
                                                      \label{eq-state-defD}
\end{align}
and the dynamics of the system is captured in the equation
\begin{align}
\dot{\pmb{x}}_D(t) ~=~ \begin{bmatrix} \dot{x}_D(t) \\
                       \dot{v}_D(t) \end{bmatrix}
                   ~=~ \pmb{f}(\dot{x}_D(t),u_D(t))
                   ~=~ \begin{bmatrix} v_D(t) \\ u_D(t) \end{bmatrix}
                                                      \label{eq-state-dotD}
\end{align}
where
\begin{align}
u_D(t) ~=~ \dot{v}_D(t) ~=~ - k a_D T                   \label{eq-cont-uD}
\end{align}
as given by (\ref{eq-dot-vD}), is the external control.
We also have the initial condition
\begin{align}
\pmb{x}_D(0) ~=~ \begin{bmatrix} 0 \\ V \end{bmatrix} \label{eq-state-initD}
\end{align}

It is hard not to notice how much more complicated the above is compared
with (\ref{eq-v-dot=-a})-(\ref{eq-state-init}). It is always so with discrete
systems --- hence the strong desire to model systems in the continuous
domain. The very rapid ramp-up in complexity when we consider the discrete
version of a continuous problem is our justification for restricting to a
particularly simple example. The ability to keep the complexity still low
enough to permit an exact solution, is extremely useful in an investigation
such as this one, allowing a comparison between exact and approximate
approaches to be made with confidence.

\begin{figure}[t]
\begin{center}
\includegraphics[width=0.55\linewidth,
                 keepaspectratio=true,viewport=10 200 440 410,
                 clip=true]
{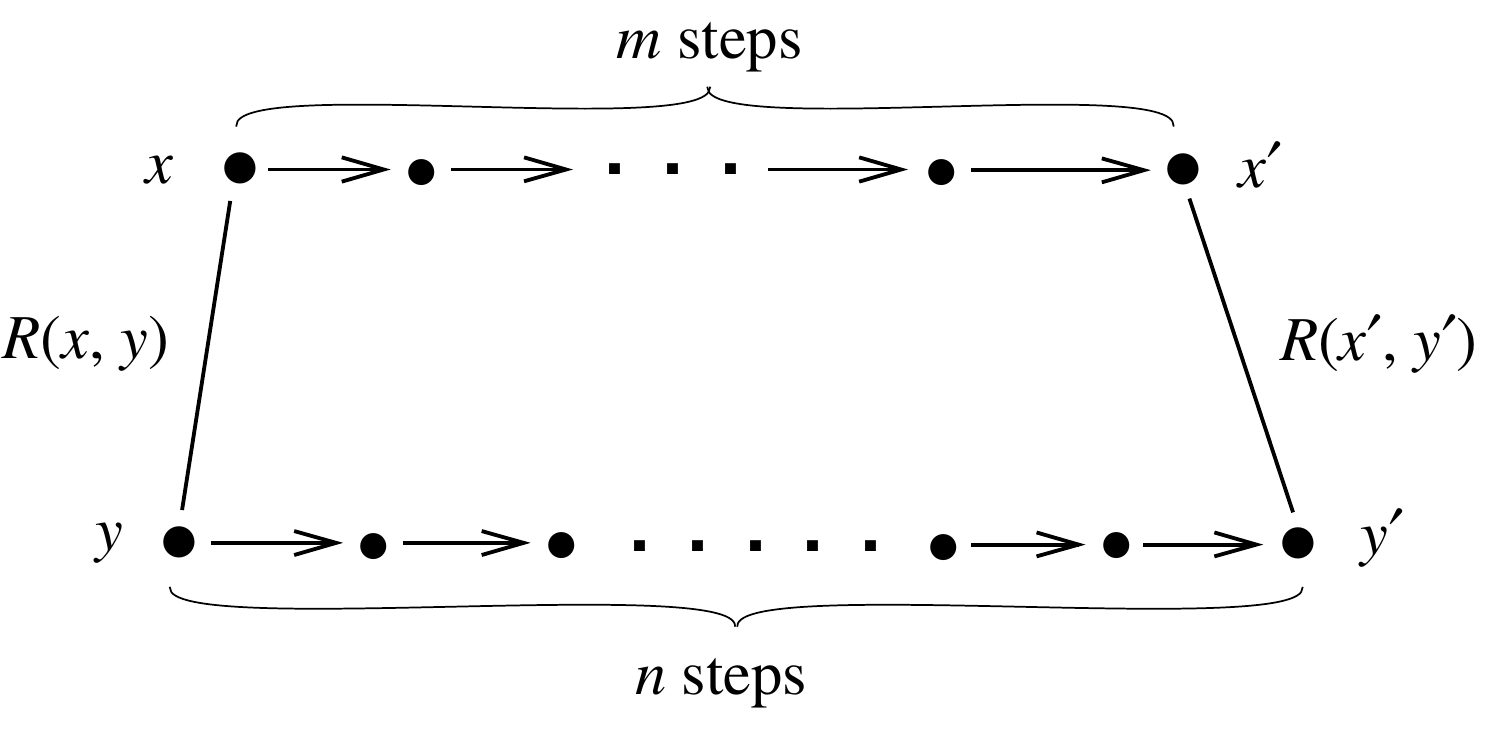}
\end{center}
\caption{An ASM $(m,n)$ diagram, showing how $m$ abstract steps, going
from state $x$ to state $x'$ simulate $n$ concrete steps, going from
$y$ to $y'$. The simulation is embodied in the retrieve relation $R$,
which holds for the before-states of the series of steps $R(x,y)$, and is
re-established for the after-states of the series $R(x',y')$.}
\label{fig-asm-m-n}
\end{figure}

\section{ASM Refinement and Retrenchment}
\label{sec-ASMRefRet}

In this section we review what we need of ASM refinement and retrenchment,
which will be the vehicles for formalization in this paper.
The standard reference for the ASM method is \cite{ASM}, building on the
earlier \cite{Boerger-ASMref03}. In general, to prove an ASM refinement,
one verifies so-called $(m,n)$ diagrams, in which $m$ abstract steps
simulate $n$ concrete ones.
The situation is illustrated in Fig.~\ref{fig-asm-m-n},
in whch we suppress input and output for clarity. For this paper, it will
be sufficient to focus on the refinement proof obligations (POs) which are
the embodiment of this policy. The first is the initialization PO:
\begin{align}
& \forall\, y' \bullet CInit(y') \Rightarrow
      (\exists\, x' \bullet AInit(x') \land R(x',y'))        \label{eq-init}
\end{align}
In (\ref{eq-init}), it is demanded that for each concrete initial state
$y'$, there is an abstract initial state $x'$ such that the retrieve or
abstraction relation $R(x',y')$ holds.

The second PO is correctness, and is concerned with the verification of
the $(m,n)$ diagrams. For this, we have to have some way of deciding
which $(m,n)$ diagrams are sufficient for the application. Let us assume
that we have done this. Let $CFrags$ be the set of fragments of concrete
execution sequences that we have previously determined will permit a
covering of all the concrete execution sequences of interest for the
application. We write
$y::ys::y' \in CFrags$ to denote an element of $CFrags$ starting with
concrete state $y$, ending with concrete state $y'$, and with intervening
concrete state sequence $ys$. Likewise $x::xs::x' \in AFrags$ for abstract
fragments. Also, let $is,js,os,ps$ denote the sequences of abstract inputs,
concrete inputs, abstract outputs, concrete outputs, respectively, belonging
to $x::xs::x'$ and $y::ys::y'$, and let $In(is,js)$ and $Out(os,ps)$ denote
suitable input and output relations. Then the correctness PO reads:
\begin{align}
& \forall\, x,is,y,ys,y',js,ps \bullet y::ys::y' \in CFrags \land
    R(x,y) \land In_{AOps,COps}(is,js)~\land                       \notag \\
& \quad COps(y::ys::y',js,ps)  \notag \\
& \qquad \Rightarrow (\exists\, xs,x',os
    \bullet AOps(x::xs::x',is,os) \land R(x',y') \land
            Out_{AOps,COps}(os,ps))                         \label{eq-refcorr}
\end{align}
In (\ref{eq-refcorr}), it is demanded that when there is a concrete
execution fragment of the form $COps(y::ys::y',js,ps)$, carried out by a
sequence of concrete operations $COps$, with state sequence $y::ys::y'$,
input sequence $js$ and output sequence $ps$, such that the retrieve and
input relations $R(x,y) \land In(is,js)$ hold between concrete and abstract
before-states and inputs, then an abstract execution fragment
$AOps(x::xs::x',is,os)$ can be found to re-establish the retrieve and output
relations $R(x',y') \land Out(os,ps)$.

The ASM refinement policy also demands that
non-termination be preserved from concrete to abstract, but we will not
need that in this paper. We now turn to retrenchment.

For retrenchment, \cite{BPJS:07a,BJP:08} give definitive accounts;
latest developments are found in \cite{RET}. See also \cite{Ban:09a}
for formulations of retrenchment adapted to several specific model
based refinement formalisms including ASM. Like refinement, retrenchment
is also characterized by POs: an initialization PO identical to
(\ref{eq-init}), and a ``correctness'' PO which weakens (\ref{eq-refcorr})
by inserting \textit{within},
\textit{output} and \textit{concedes} relations, $W_{Op},O_{Op},C_{Op}$
respectively into (\ref{eq-refcorr}), to give extra flexibility and
expressivity. In particular, the concession $C_{Op}$ weakens the conclusions
of (\ref{eq-refcorr}) disjunctively, giving room for many kinds of
``exceptional'' behaviour. The result is:
\begin{align}
& \forall\, x,is,y,ys,y',js,ps \bullet y::ys::y' \in CFrags \land
    R(x,y) \land W_{AOps,COps}(is,js,x,y)~\land                   \notag \\
& \quad COps(y::ys::y',js,ps) \notag \\
& \qquad \Rightarrow (\exists\, xs,x',os \bullet AOps(x::xs::x',is,os)~\land
                                                                  \notag \\
& \qquad\qquad\qquad\quad ((R(x',y') \land
        O_{AOps,COps}(x::xs::x',is,os,y::ys::y',js,ps))~\lor      \notag \\
& \qquad\qquad\qquad\quad~~~C_{AOps,COps}(x::xs::x',is,os,y::ys::y',js,ps)))
                                                          \label{eq-retcorr}
\end{align}
To ensure that retrenchment only deals with well defined transitions, and
to ensure smooth retrenchment/refinement interworking, we also insist that
$R \land W_{Op}$ always falls in the domain of the requisite operations,
though this is another thing not needed here.

\begin{figure}[t]
\begin{center}
\includegraphics[width=0.35\linewidth,
                keepaspectratio=true,viewport=35 730 180 830,
                clip=true]
{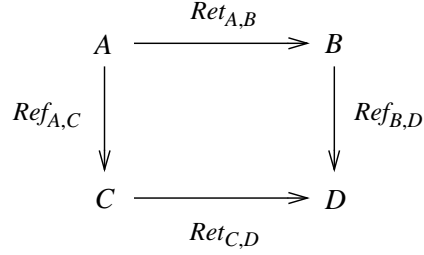}
\end{center}
\caption{The \textit{Tower Pattern} basic square, with
refinements vertical, retrenchments horizontal.}
\label{fig-TP}
\end{figure}

The smooth interworking between refinements and retrenchments is guaranteed
by the \textit{Tower Pattern}.
The basic construction for this is shown in Fig.~\ref{fig-TP}.
There, refinements are vertical arrows and retrenchments
are horizontal, and the two paths round the square from $A$ to $D$ (given
by composing $Ref_{A,C}$ with $Ret_{C,D}$ on the one hand, and on the other,
by composing $Ret_{A,B}$ with $Ref_{B,D}$) are compatible, in the sense that
they each define a portion of a (potentially larger) retrenchment from $A$
to $D$.

At this point one might legitimately ask what all the above has to do with
our case study, in which the dynamics that we considered is entirely in
the continuous domain (albeit taking into account discontinuous control
inputs when necessary). The answer lies in the focus on the use of paths
through the system at both abstract and concrete levels in the POs of ASM.
With this focus, it is unproblematic to reconfigure the $(m,n)$ rules
(\ref{eq-refcorr}) and (\ref{eq-retcorr}) to deal with continuous paths
rather than discrete ones. Thus $CFrags$ and $AFrags$ can now refer to
fragements of continuous system trajectories, rather than sequences of
state-to-state hops. Likewise the $is$ and $js$ in $W_{AOps,COps}(is,js,x,y)$
now refer to the continuous input signals along the trajectories, and
so on for the other terms in (\ref{eq-refcorr}) and (\ref{eq-retcorr}).
We see this exemplified in detail in the retrenchment of Section 
\ref{ssec-RigBoundsRet}.

\section{Formalizing the Continuous to Discrete Modeling Change}
\label{sec-FormCtDisc}

In the control literature, one finds many ways of discretizing continuous
designs (see loc.~cit.), and the evaluation of the relationship between
continuous and discrete is often based on \textit{ad hoc} engineering
rules of thumb. While these typically yield perfectly good results in
practice, the criteria used fall far short of the kind of precision
needed for a good fit with model based formal development techniques.
As a consequence, when model based formal development techniques are used
to support the digital implementation of the discrete counterpart of some
continuous design, the formal modeling inevitably starts already in the
discrete domain. Obviously this yields a weaker formal support for the
process than if the formal modeling had started earlier, at the continuous
design stage, and was integrated into all the subsequent design steps,
including the change from continuous to discrete.

Our objective in this paper is to illustrate how to make a judgement
about the discretization of a control problem, that has enough precision
to integrate well with model based formal technologies. To achieve
this we have recourse to the rigorous theory of ODEs. It can be
shown\footnote{In the extended version of this paper it \textit{is} shown.}
that two instances of a control problem which differ solely in the input
control satisfy an inequality:
\begin{align}
||\pmb{x}^u-\pmb{x}_D^{u_D}|| &~\leq~ K\!\mathit{2} ||u - u_D||_2
                                                \label{eq-K2-u-uD}
\end{align}
In (\ref{eq-K2-u-uD}), $||\pmb{x}^u-\pmb{x}_D^{u_D}||$ is the
$\mathcal{L}^\infty$ norm of $\pmb{x}^u-\pmb{x}_D^{u_D}$, or,
in plain English, the maximum value over the interval
$[0 \ldots T_{Stop}]$ attained by the difference between
continuous and discrete values of any state component.
Likewise, $||u - u_D||_2$ is the $\mathcal{L}^2$ norm of $u - u_D$,
or, in plain English, the root integrated square difference between $u$ and
$u_D$, calculated over the interval $[0 \ldots T_{Stop}]$. Finally,
$K\!\mathit{2}$ is a constant.

We note that the continuous and
discrete versions of our case study, with initial states
(\ref{eq-state-init}) and (\ref{eq-state-initD}), over the time
interval from $0$ to $T_{Stop}$, characterize just such a scenario,
since (\ref{eq-state-dot}) and (\ref{eq-state-init}) differ from
(\ref{eq-state-dotD}) and (\ref{eq-state-initD}) only in the use of
$u_D$ rather than $u$ among the independent variables.

\subsection{Rigorous Bounds on Continuous and Discrete Systems}
\label{ssec-RigBounds}

We now flesh out what (\ref{eq-K2-u-uD}) means for our little case study.
We consider the values of the quantities on the right hand side
of (\ref{eq-K2-u-uD}) in order to obtain a bound for the value of the
left hand side. Referring to (\ref{eq-K2-u-uD}), theory furnishes an
explicit value for the constant $K\!\mathit{2}$, namely
\begin{align}
K\!\mathit{2} &~=~ e^{K_{\pmb{f}}} ||k_u||_2
                                                   \label{eq-K2=KfKfu}
\end{align}
In (\ref{eq-K2=KfKfu}) $K_f$ is $k_f\,T_{Stop}$, where $k_f$ is the
$\mathcal{L}^\infty$ norm of $\pmb{f}_{\pmb{x}}$, or, the absolute
maximum value (over the interval $[0 \ldots T_{Stop}]$) of the Lipschitz
constant governing the variation of the control law $\pmb{f}$ with respect
to the state. In our application, the form of the control law is
\begin{align}
\pmb{f}(v(t),u(t)) ~=~ \begin{bmatrix} v(t) \\ u(t) \end{bmatrix}
                                                  \label{eq-f-form}
\end{align}
and it is clear that there is only one component of $\pmb{f}$ with
a non-zero partial derivative with respect to either $x$ or $v$,
namely the first
\begin{align}
\frac{\partial \pmb{f}_1}{\partial v} ~=~ 1     \label{eq-pd-of-f-v}
\end{align}
With this, the first factor of (\ref{eq-K2=KfKfu}) is just $e^{T_{Stop}}$.

Regarding the second factor, $||k_u||_2$ is the root integrated square
value of the Lipschitz constant governing the variation of the control
law with respect to the input control signal. Again there is only one
component of $\pmb{f}$ with a non-zero partial derivative with respect
to $u$, namely the second
\begin{align}
\frac{\partial \pmb{f}_2}{\partial u} ~=~ 1     \label{eq-pd-of-f-u}
\end{align}
so the root integrated square reduces to $\sqrt{T_{Stop}}$. So we get
\begin{align}
K\!\mathit{2} &~=~ e^{T_{Stop}} \sqrt{T_{Stop}}             \label{eq-K2-num}
\end{align}

Turning to the second factor on the right hand side of (\ref{eq-K2-u-uD}),
$||u - u_D||_2$, we recall that we know explicitly what $u$ and $u_D$ are
from our earlier calculations. From (\ref{eq-cont-u}) and (\ref{eq-cont-uD})
we know that
\begin{align}
u(t) &~=~ -a t & u_D(t) ~=~ - k a_D T                 \label{eq-cont-u-again}
\end{align}
where, from (\ref{eq-V=aT}) and (\ref{eq-VD})
\begin{align}
a &~=~ \frac{2V}{T_{Stop}^2} & a_D &~=~ \frac{2V}{T_{Stop}^2(1+1/N)}
                                                        \label{eq-V=aT-again}
\end{align}
Now (\ref{eq-cont-u-again}) shows that $u(t)$ decreases linearly,
and that $u_D(t)$ is a staircase
function, decreasing in equal sized steps near $u(t)$. It is
clear from (\ref{eq-cont-u-again}) that in the limit $t \rightarrow 0+$,
we have $u(0+)=0$ and $u_D(0+)=-a_D T$, so that $u(0+)-u_D(0+) = a_D T$.
It is also clear from (\ref{eq-cont-u-again}) that in the limit
$t \rightarrow T_{Stop}-$, we have $u(T_{Stop}-) = - a T_{Stop}$ and
$u_D(T_{Stop}-) = - N a_D T = - a_D T_{Stop}$, so that
$u(T_{Stop}-)-u_D(T_{Stop}-) = (a_D-a)T_{Stop}
= a_DT_{Stop}[1 - (1+1/N)] = -a_DT_{Stop}/N = -a_DT$.
Since the staircase has equal sized steps, it evidently the case that
the staircase $u_D(t)$ ranges around $u(t)$ within a bound $a_DT$.
\begin{align}
|u(t) - u_D(t)| ~\leq~ a_D T                     \label{eq-staircase-bound}
\end{align}
This furnishes a suitable overestimate for the root integrated square
difference between $u(t)$ and $u_D(t)$ as follows
\begin{align}
||u - u_D||_2 ~\leq~ \sqrt{\int_{t=0}^{T_{Stop}} [a_DT]^2 dt} ~=~
        a_D T \sqrt{T_{Stop}}                                \label{eq-rms}
\end{align}
Substituting all the values we have obtained into (\ref{eq-K2-u-uD}), we get
\begin{align}
||\pmb{x}^u-\pmb{x}_D^{u_D}|| ~\leq~
        e^{T_{Stop}} \sqrt{T_{Stop}} \times a_D T \sqrt{T_{Stop}}
        ~=~ e^{T_{Stop}} a_D T T_{Stop}
                                                \label{eq-explicit}
\end{align}
We see that despite the potential for the deviation between $u(t)$
and $u_D(T)$ to grow exponentially with the size of the time interval,
a possibility severely exacerbated by our rather crude bound (\ref{eq-rms}),
it is always possible to reduce it by an arbitrary amount by making
the discretization, measured by $N$, fine enough.

\subsection{Turning Rigorous Bounds into Retrenchment Data}
\label{ssec-RigBoundsRet}

Now that we have a precise relationship between the continuous and
discrete control systems, we can look to incorporate this into our
model based formal description.

In general, the exigencies of model based formal refinement are too
exacting to be able to accommodate the kind of relationships just derived.
Retrenchment though, has been purposely designed to be more forgiving
in this regard, so that is what we will use.

Regardless though, of which model based formal description technique
is adopted, is the issue that all such techniques are designed for discrete
state transitions, and presume a well defined notion of ``next state'',
to which an equally clear notion of ``current state'' can be related.

In continuous dynamics there is no sensible notion of next state that we can
immediately use. However, as we noted above, the $(m,n)$ diagram approach of
ASM refinement makes clear that it is \textit{paths} at abstract and concrete
levels that are being related. Thus, although we avoid technical details in
this paper, we extend the ASM approach to incorporate \textit{continuous paths}
as well as discrete ones. The incentive to do this was one strong reason for
choosing ASMs in this work. (Note that this perspective on refinememt between
paths is equally applicable to both the continuous and discretized versions of
our control problem. In the continuous problem there is a single continuous
path. In the discretized problem there are $N$ consecutive shorter continuous
``zero order held'' paths, interleaved, at the instants at $kT$, by the discrete
recalculations of the output signal, thus constituting a path comprising both
continuous and discrete components.)

Since the rigorous results we use concern the same starting state for the
two systems, our formal statement is constrained to be an end-to-end one.
It will express an end-to-end relationship between the smooth dynamics
at the continuous level, and the discretized level's dynamics (which is
continuous too, though punctuated at every multiple of $T$ by a discontinuous
change in the acceleration).

As we saw before,
a retrenchment between two specific operation sequences consists of
four things: a retrieve relation between the state spaces, a within
relation for the before-states and inputs, an output relation for
the after-states and outputs (and before-states and inputs too if
necessary), and a concedes relation for the after-states and outputs
(and before-states and inputs too if needed). In the relations below,
we use some \textit{ad hoc} notations whose meaning should be obvious
from the preceding material.

Regarding the retrieve relation $R$, there is a very natural one that
we might expect to use, namely the identity between state values in
the continuous and discretized worlds. However, even though in our
specific case study the two models start out in the same state thus
making such a putative $R$ true in the hypothesis of the PO
(\ref{eq-retcorr}), in most cases, that assumption will not hold, and
so we prefer to follow a more generic approach, which will be applicable
in a wider set of scenarios. A second proposal for $R$ would see it
express a margin of tolerance between the state values in the continuous
and discretized worlds, as discussed in Section \ref{sec-RelWork}. This
proposal would also work after a fashion, but such a proposal works best
when the relationship between the two system states is stable throughout
the dynamics --- we have then a kind of refinement. In our case study,
this assumption does not hold since the discrepancy between the two system
states grows steadily through the dynamics.

To accomodate inconvenient situations such as these, retrenchment makes
provisions for expressing the relationship (or just aspects of the relationship)
between the states at the before- point of the transition being discussed
in the within relation $W$ instead of (or in addition to) in $R$. Since
the facts expressed in $W$ do not need to be re-established in the conclusion
of the PO (\ref{eq-retcorr}), this provides the most flexible way of
incorporating appropriate facts about the systems' before-states in the PO.
With this strategy, a global retrieve relation is not appropriate, and we set
$R$ to \textsf{true}
\begin{align}
R(\langle x(t),v(t) \rangle,\langle x_D(t),v_D(t) \rangle)
  ~\equiv~ \textsf{true}                       \label{eq-our-R}
\end{align}

The job of expressing that the before-states are suitably matched in the PO,
taking into account the input control signals throughout the interval of interest,
is thus taken on by the within relation $W$
\begin{align}
& W(u(t \in [0 \ldots T_{Stop}]),u_D([t \in 0 \ldots T_{Stop}]),
    \langle x(0),v(0) \rangle,\langle x_D(0),v_D(0) \rangle)
         ~\equiv~                                                    \notag \\
& \quad x(0) = x_D(0)
  \land v(0) = v_D(0)~~~~\land~~~~||u - u_D||_2 \leq a_D T \sqrt{T_{Stop}}
                                                              \label{eq-our-W}
\end{align}
Note that while $W$ relates just the continuous and discrete before-states,
it also relates the whole of the continuous and discrete control inputs.

The output relation $O$ says what happens at the end of the period of
interest. In our case, on the basis of the rather heavy calculations
that came earlier, we can use $O$ to say that the after-states diverge
by no more than the bound derived in (\ref{eq-explicit})
\begin{align}
& O(\langle x(T_{Stop}),v(T_{Stop}) \rangle,
    \langle x_D(T_{Stop}),v_D(T_{Stop}) \rangle)
         ~\equiv~                                                    \notag \\
& \quad |x(T_{Stop}) - x_D(T_{Stop})| \leq
  e^{T_{Stop}} a_D T T_{Stop}~~~~\land~~~~|v(T_{Stop}) - v_D(T_{Stop})| \leq
  e^{T_{Stop}} a_D T T_{Stop}                                 \label{eq-our-O}
\end{align}
Note that although $O$ itself speaks \textit{explicitly} only about the
after-states that are attained by the two systems, the fact that we derived
the properties of the after-states in question using an $\mathcal{L}^\infty$
analysis, means that the same bound holds \textit{throughout} the interval
of interest. The advantage of this formulation is that we automatically get
a discreteness of the description in terms of before- and after- states, which
will integrate neatly with discrete system reasoners (in the event that such
modeling is eventually incorporated into mechanised tools), while yet providing
guarantees that hold throughout the interval of interest.

Since our system is so simple, $O$ already captures all that we need
to say, and the kind of exceptional behaviour that may need to be taken
into account in more realistic engineering situations is not present.
This is also connected wsiyth the fact that we have trivialised the
retrieve relation. Accordingly we can set the concedes relation $C$
to \textsf{false}
\begin{align}
& C(\langle x(T_{Stop}),v(T_{Stop}) \rangle,
    \langle x_D(T_{Stop}),v_D(T_{Stop}) \rangle)
         ~\equiv~ \textsf{false}                    \label{eq-our-C}
\end{align}

With these data, the proof obligation (\ref{eq-retcorr}) becomes provable
on the basis of the results cited earlier, which establishes the formal
connection between the continuous and discrete domains in a way that can
be integrated with formal refinements on both the continuous and discrete
sides.

Particularly noteworthy is the fact that the discrepancy between the
states grows linearly with time; and that this is a property of the exact
solutions and not just an artifact of some approximation scheme. If we
tried to handle this in a pure refinement framework, using a retrieve
relation $R$ to capture the relationship between states in the two
models (regardless of whether $R$ was an exact, pointwise relationship,
or an approximate one, analogous to the approximate simulation relations
discussed in Section \ref{sec-RelWork}), then assuming such an $R$ for
the before-states would not enable us to re-establish it for the
after-states, and the correctness PO could not be proved. The greater
flexibiity of retrenchment permits us to handle the before-states in
the within relation and the after-states in the output relation,
overcoming the problem.

\subsection{Corroboration}
\label{ssec-Corrob}

In our case study, exact solvability of the control models in both
continuous and discrete domains gives us additional and independent
confirmation of the approach we are advocating in this paper.

Both continuous and discrete models ``run'' for the same amount of time,
$T_{Stop}$, and the output relation (\ref{eq-our-O}) gives an estimate for
the discrepancy between the continuous and discrete states reached in the
two models after that time. The states themselves consist of two components,
the displacements and the velocities.

Regarding the velocities, both models come to a standstill after exactly
$T_{Stop}$. Consequently both $v(T_{Stop})$ and $v_D(T_{Stop})$ are zero,
so that $|v(T_{Stop}) - v_D(T_{Stop})| = 0$, and any positive upper bound
is bound to be sound. So (\ref{eq-our-O}), which gives the overestimate
$e^{T_{Stop}} a_D T T_{Stop}$ for $|v(T_{Stop}) - v_D(T_{Stop})|$ is correct
regarding the velocities, but in an unsurprising way. 

Regarding the displacements, the quantization of $T_{Stop}$ in the discrete
case, leads to the continuous and discrete dynamics stopping at slightly
different places, $D$ and $D_D$ respectively, which we calculated earlier.
On that basis, we can calculate the exact difference (disregarding
$O(N^{-2})$ and beyond):
\begin{align}
|x(T_{Stop}) - x_D(T_{Stop})| &~=~ 
      \frac{2}{3} \frac{V T_{Stop}}{4N} ~=~
      \frac{1}{2} a_D T_{Stop}^2
           \left(1 + \frac{1}{N} \right) \frac{T_{Stop}}{6N}       \notag \\
   &~=~ \frac{1}{12} a_D T T_{Stop}^2 \left(1 + \frac{1}{N} \right)
                                                       \label{eq-disp-exact}
\end{align}
On the other hand, the output relation (\ref{eq-our-O}) gives the estimate
$e^{T_{Stop}} a_D T T_{Stop}$ for this quantity. Thus the exact value falls
within the bounds of the estimate, as it should, if and only if (after
cancelling the common factor $a_D T T_{Stop}$):
\begin{align}
& \frac{T_{Stop}}{12} \left(1 + \frac{1}{N} \right) ~\leq~ e^{T_{Stop}}
                                                       \label{eq-disp-Corr}
\end{align}
Since a linear function of $T_{Stop}$ of slope less than $1$ can never catch
an exponential function of $T_{Stop}$ with coefficient $1$,
(\ref{eq-disp-Corr}) is obviously true, and we have our corroboration.

\section{Continuous to Discrete Modeling in a Wider Design Process}
\label{sec-ContDes}

The previous sections focused in detail on how the rigorous theory of
ODEs was capable of yielding results that could be integrated with
existing model based refinement centred development methodologies,
all in the context of a very simple example. The essence of the
process is to identify useful results from the mathematical theory,
and then to drill down into the details of the proof to identify explicit
values for the constants etc.~that figure in them. The latter process
is often required, since it is frequently the case that the goal of a proof
of interest is satisfied by merely asserting the existence of the requisite
constant, without a specific value being calculated, since that is usually
enough to enable the existence of some limit to be proved. By contrast,
for us, the existence of the limit is insufficient, since no engineering
process can completely traverse the infinite road required to reach it.
Rather, we need the explicit value of everything, so that we can judge
how far down the road we have to go before we can be sure that we have
gone ``far enough'' to achieve the engineering quality we require.

In this section, we outline how a retrenchment obtained in this way could
be placed in the context of a development methodology of wider scope. For
lack of space we touch on a number of technical issues that are only
dealt with properly in the extended version of this paper.
The key idea for the integration is the Tower Pattern, mentioned already
in Section \ref{sec-ASMRefRet}. This allows the extreme flexibility
of retrenchment with its ability to accomodate a very wide variety
of system properties, to be shored up with the much stricter guarantees
that model based refinement offers, the latter coming at the price
of much more restricted expressivity as regards system properties.
Although we do not have the space to discuss the point at length,
we claim that a judicious combination of the two techniques can give
better coverage of the route from high level domain centred requirements
goals to low level implementation, than either technique alone. Thus
on the one hand, use of refinement alone, forces the consideration of
and commitment to, low level restrictions such as finiteness limits on
arithmetic, far too early in the process, in order that all later
models can (in effect) be conservative extensions of their predecessors.
On the other hand, use of retrenchment alone makes it much harder to
track how system properties evolve as the development proceeds, since
successive models can be connected to their predecessors in a very
loose manner, requiring much tighter focus on \textit{post hoc} validation.

In our case, it is appropriate to use retrenchment to capture the properties
of the discretization step, since that is something that has eluded model
based refinement techniques.\footnote{It has to be noted that the introduction
of approximate simulations has improved the situation recently with regard
to stable systems, but in a more general context the observation remains true.}
However, either side of the discretization step, we are free to use refinement,
since on each side individually, the models display much more consistency
regarding the kind of properties that can be handled with sufficient eloquence
using refinement alone.

\begin{figure}[t]
\begin{center}
\includegraphics[width=0.7\linewidth,
                 keepaspectratio=true,viewport=35 230 540 520,
                 clip=true]
{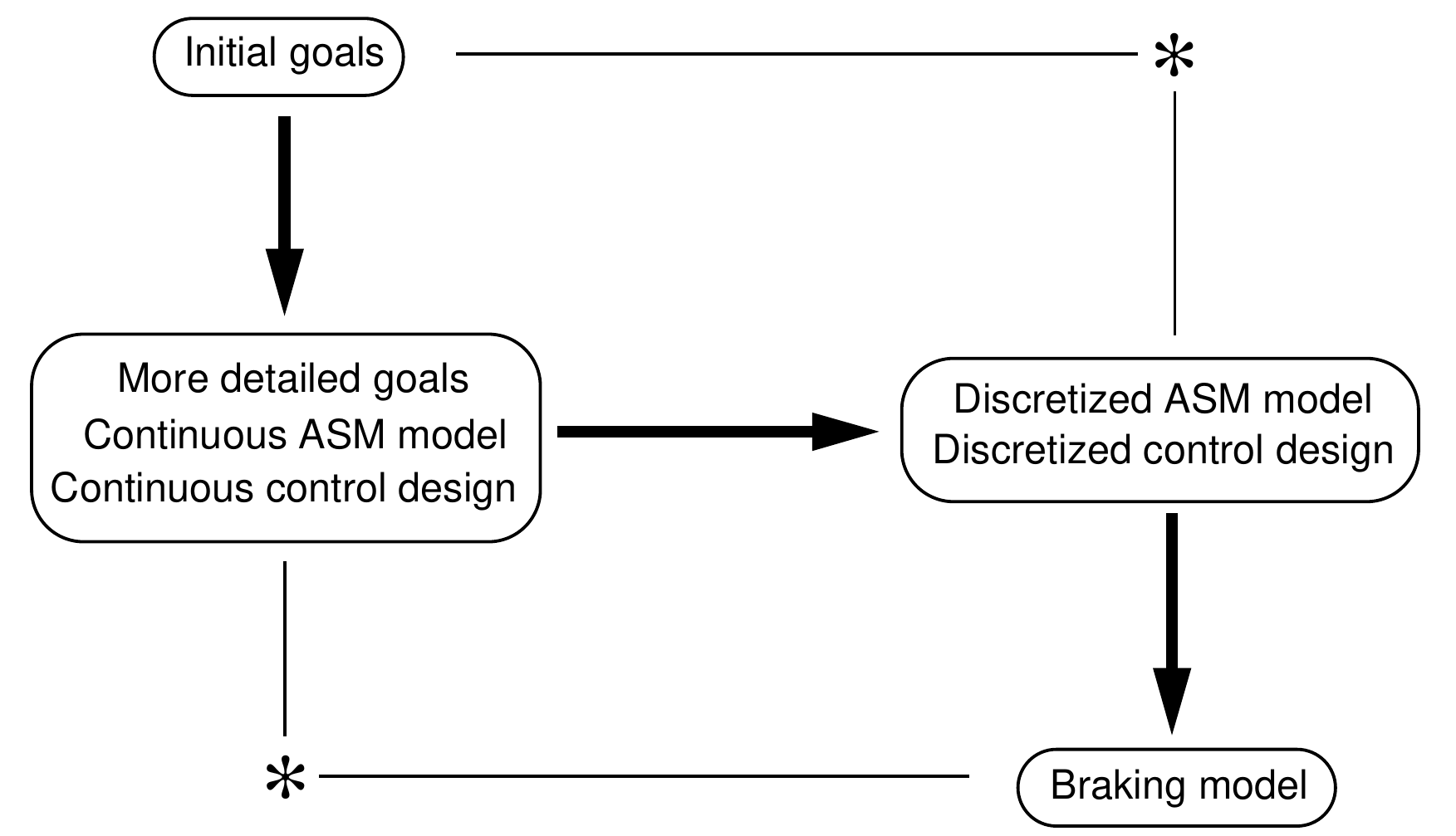}
\end{center}
\caption{An overview of a complete development, starting with
abstract goals, proceeding through explicit continuous and discrete
deceleration models, and continuing with further low level models. Vertical
arrows are (perhaps successive) model based refinements. Horizontal arrows
are retrenchments, suited to relating models too different to be connected
by refinement.}
\label{fig-whole}
\end{figure}

The complete process that we have in mind may be summarized in
Fig.~\ref{fig-whole}. The thick arrows trace a path through
a family of models that a development route could plausibly take.
The left hand side of the diagram concerns continuous models. 
At the start, we have high level requirements goals, expressed
in a notation with formal underpinnings. We have in mind a formalism
like KAOS \cite{KAOS,Let:01} (or more precisely, an adaptation of it
to deal more honestly with continuous processes). These requirements goals
can then be formally refined till they can be \textit{operationalized},
i.e.~transformed into the operations of a methodology such as ASM
(again, adapted to deal with continuous evolution). Then comes
our discretization step, necessitating the use of retrenchment.
Once we have crossed the continuous/discrete boundary, we are free
to revert to traditional model based refinement techniques for
discrete state transition systems --- no worries about continuous
phenomena any more. In Fig.~\ref{fig-whole} we indicate how the discrete
kinematics that we investigated earlier might be refined to a model
of train braking, in which concern with the dynamics is replaced by
a focus on the actuators that would implement the deceleration increments
in practice.

Fig.~\ref{fig-whole} also features other models, indicated by asterisks.
These are models whose existence is guaranteed by the Tower theorems
\cite{BaJe:09,Jes:05}, making the squares of Fig.~\ref{fig-whole}
commuting in an appropriate sense. However, we argue that these models
are less useful than the others. Thus the lower left model would be a
continuous version of the braking model, an unrealistic overidealisation
so close to implementation. The upper right model would be a discretized
version of the highest level requirements goals for train stopping. Again
this would be inappropriate at such a high level, since it clutters what
ought to be the most perspicuous expression of the system goals with a
lot of material concerning low level details of the discretization scheme.
This bears out what we said above about a combination of refinement and
retrenchment techniques providing the best coverage of the route from high
level requirements to low level implementation.

Above, we mentioned adaptations of KAOS and ASM to deal with continuous 
behaviour. We discuss these briefly now. Regarding ASM, a major part of what
we need is already available in the literature, eg.~\cite{CoSl:08,SlVa:08}
which deal with \textit{(Real) Timed ASM}. The essential observation is
that in the context of continuous time, system states should be modeled
as persisting over half-open half-closed time intervals, eg.~$(t_0,t_1]$.
This allows the typical discontinuous state transition in a typical discrete
transition system, say of a state variable $v$, to be represented as the
move from $v(t_0)$ (the value of $v$ at $t_0$, which lies outside $(t_0,t_1]$
and is the right hand endpoint of the preceding interval), to
$\lim_{\epsilon \rightarrow 0+} v(t_0+\epsilon)$ (the left hand limit at
$t_0$ from the right, of values of $v$ within the interval $(t_0,t_1]$).
Likewise, a period of continuous evolution can be understood as persisting
over such a half-closed interval, governed by a suitably well posed ODE initial
value problem, and with the truth of the initial conditions for the initial
value problem at the end of the preceding interval being the trigger for the
system's subsequently following a trajectory specified by the ODE problem.
With these conventions, a version of ASM in which discrete steps alternate
with continuous flows can be developed, reflecting many of the characteristics
of hybrid automata.

A similar approach can be adopted for KAOS. Although KAOS depends on a notion
of time from the outset, in the normal KAOS formalism, time is discrete,
typically indexed by the integers, with requirements goals expressed as
temporal logic formulae over time. For a version over continuous time,
while some temporal operators, eg.~\textit{always}, \textit{until}, offer
no conceptual difficulties, the \textit{next} operator needs to be rethought.
Again half-open half-closed intervals, with successor states being defined
via the limit from the right at the left hand end of a half-closed
interval, can be used. To avoid problems arising due to an accumulation of
\textit{next} operators, syntactic restrictions have to be imposed on the
permitted temporal formulae. However, the kinds of restrictions that need
to be imposed are satisfied by the patterns that KAOS requirements are
normally built out of.

\section{Conclusion}
\label{sec-concl}

In this paper we introduced a small continuous control problem in state space
format, and then treated a discretized counterpart of it, utilising a zero
order hold. Then came the main novel contribution of the paper, a rigorous
treatment of the continuous to discrete modeling transformation, based on
cited results from ODE theory. That done, we were able to integrate the
results into a retrenchment which related from continuous and discrete
models. As noted earlier, model based formal development normally starts
already in the discrete domain, so the ability to connect this with the
continuous world in a reasoned way, is a significant extension of the
potential of model based formal techniques to underpin developments of
such systems. Equally importantly, in making essential use of retrenchment
to forge the connection between continuous modeling and discrete modeling,
this work gives a fresh confirmation of the utility of the concept as a
worthwhile adjunct to refinement in tackling the wider issues
connected with real world formal developments.

Of course, this paper is by no means the last word in developments
of this kind. As well as tackling a control problem that was almost
trivial technically, the rigorous result from mathematical control
theory that we utilized was relatively limited, insisting, as it did,
that the two behaviours that were compared, started from the same
state, using a rather crude $\mathcal{L}^2$ estimate of the difference
in the control inputs to derive its conclusion, and being based on
rather generic properties of the ODEs that govern the dynamics of
the control problem. (These simple contraints also meant that relatively
little of the expressive power of retrenchment was used in this case study.)
In more realistic cases, the problem will be less amenable to analytic
solution, and feedback mechanisms will help alleviate the inherent
uncertainty that arises. Moreover, while a crude $\mathcal{L}^2$ estimate
of the difference in the control inputs allows the two control inputs to
get as far away from each other as the bounds on the control space allow,
in practice, feedback mechanisms will tend to push them together, and this
could be exploited to derive more stringent estimates of the difference
between continuous and discrete control. All of this remains to be discussed
in future work, as does the extension of the KAOS and ASM formalisms (or
any alternatives that might be contemplated to act in their place), that
can encompass the continuous behaviours that we have described.

Our work is to be contrasted with the possibilities offerd by the hybrid
systems approach \cite{Tab:09}. There, the insistence on (approximate)
bisimulation between a continuous system and a discrete counterpart
restricts attention to control systems which are stable in the Liapunov
sense. In any event, the intense focus on considerations of algorithmic
decidability in that field, with automata homomorphism as such a prominent
relationship between system models, can inhibit design expressivity for the
purposes that concern us. For instance, techniques that rely on stability,
are, strictly speaking, not applicable to our simple case study.

Once a suitable collection of widely applicable and useful results of
the kind discussed here have been established, the way is open for the
incorporation of these into appropriate formal development tools. These
would be of a different flavour to those typically developed for the hybrid
systems field, since they would have more emphasis on interactive proving
than is typically the case there. One snag that would have to be overcome
is that most proving based tools cope rather badly with the kind of applied
mathematics and rigorous analysis techniques that are required for this work.
A notable exception is the PVS suite \cite{COR:95,PVS}, for which substantial
library support exists to underpin both applied mathematics and its more
rigorous counterparts, eg.~\cite{Dut:96}. This would be the obvious jumping
off point for the development of tools that aligned well with our approach.

\bibliographystyle{eptcs}
\bibliography{rigorous.control}

\begin{thebibliography}{10}
\providecommand{\bibitemdeclare}[2]{}
\providecommand{\urlprefix}{Available at }
\providecommand{\url}[1]{\texttt{#1}}
\providecommand{\href}[2]{\texttt{#2}}
\providecommand{\urlalt}[2]{\href{#1}{#2}}
\providecommand{\doi}[1]{doi:\urlalt{http://dx.doi.org/#1}{#1}}
\providecommand{\bibinfo}[2]{#2}

\bibitemdeclare{book}{abrial:96}
\bibitem{abrial:96}
\bibinfo{author}{J-R. Abrial} (\bibinfo{year}{1996}):
  \emph{\bibinfo{title}{{The {B}-Book: Assigning Programs to Meanings}}}.
\newblock \bibinfo{publisher}{Cambridge University Press},
  \doi{10.1017/CBO9780511624162}.

\bibitemdeclare{book}{abrial:10}
\bibitem{abrial:10}
\bibinfo{author}{J-R. Abrial} (\bibinfo{year}{2010}):
  \emph{\bibinfo{title}{{Modeling in Event-B: System and Software
  Engineering}}}.
\newblock \bibinfo{publisher}{Cambridge University Press}.

\bibitemdeclare{book}{ahmed:06}
\bibitem{ahmed:06}
\bibinfo{author}{N.~Ahmed} (\bibinfo{year}{2006}):
  \emph{\bibinfo{title}{Dynamic Systems and Control With Applications}}.
\newblock \bibinfo{publisher}{World Scientific}.

\bibitemdeclare{inproceedings}{AlCouHenHo:93}
\bibitem{AlCouHenHo:93}
\bibinfo{author}{R.~Alur}, \bibinfo{author}{C.~Courcoubetis},
  \bibinfo{author}{T.~Henzinger} \& \bibinfo{author}{P-H. Ho}
  (\bibinfo{year}{1993}): \emph{\bibinfo{title}{{Hybrid Automata: An
  Algorithmic Approach to the Specification and Verification of Hybrid
  Systems}}}.
\newblock In: {\sl \bibinfo{booktitle}{Proc. Workshop on Theory of Hybrid
  Systems}}, {\sl \bibinfo{series}{LNCS}} \bibinfo{volume}{736},
  \bibinfo{publisher}{Springer}, pp. \bibinfo{pages}{209--229}.

\bibitemdeclare{article}{AlurDill:94}
\bibitem{AlurDill:94}
\bibinfo{author}{R.~Alur} \& \bibinfo{author}{D.~Dill} (\bibinfo{year}{1994}):
  \emph{\bibinfo{title}{{A Theory of Timed Automata}}}.
\newblock {\sl \bibinfo{journal}{Theor. Comp. Sci.}} \bibinfo{volume}{126}, pp.
  \bibinfo{pages}{183--235}, \doi{10.1016/0304-3975(94)90010-8}.

\bibitemdeclare{book}{Ants:06}
\bibitem{Ants:06}
\bibinfo{author}{P.~Antsaklis} \& \bibinfo{author}{A.~Michel}
  (\bibinfo{year}{2006}): \emph{\bibinfo{title}{Linear Systems}}.
\newblock \bibinfo{publisher}{Birkhauser}.

\bibitemdeclare{article}{Ban:09a}
\bibitem{Ban:09a}
\bibinfo{author}{R.~Banach}: \emph{\bibinfo{title}{{Model Based Refinement and
  the Design of Retrenchments.}}} \bibinfo{note}{Available from \cite{RET}.}

\bibitemdeclare{article}{BaJe:09}
\bibitem{BaJe:09}
\bibinfo{author}{R.~Banach} \& \bibinfo{author}{C.~Jeske}:
  \emph{\bibinfo{title}{{Retrenchment and Refinement Interworking: the Tower
  Theorems.}}} \bibinfo{note}{Submitted}.

\bibitemdeclare{article}{BJP:08}
\bibitem{BJP:08}
\bibinfo{author}{R.~Banach}, \bibinfo{author}{C.~Jeske} \&
  \bibinfo{author}{M.~Poppleton} (\bibinfo{year}{2008}):
  \emph{\bibinfo{title}{{C}omposition {M}echanisms for {R}etrenchment}}.
\newblock {\sl \bibinfo{journal}{J. Log. Alg. Prog.}} \bibinfo{volume}{75}, pp.
  \bibinfo{pages}{209--229}, \doi{10.1016/j.jlap.2007.11.001}.

\bibitemdeclare{article}{BPJS:07a}
\bibitem{BPJS:07a}
\bibinfo{author}{R.~Banach}, \bibinfo{author}{M.~Poppleton},
  \bibinfo{author}{C.~Jeske} \& \bibinfo{author}{S.~Stepney}
  (\bibinfo{year}{2007}): \emph{\bibinfo{title}{{E}ngineering and {T}heoretical
  {U}nderpinnings of {R}etrenchment}}.
\newblock {\sl \bibinfo{journal}{Sci. Comp. Prog.}} \bibinfo{volume}{67}, pp.
  \bibinfo{pages}{301--329}, \doi{10.1016/j.scico.2007.04.002}.

\bibitemdeclare{book}{barnett:75}
\bibitem{barnett:75}
\bibinfo{author}{S.~Barnett} (\bibinfo{year}{1975}):
  \emph{\bibinfo{title}{Introduction to Mathematical Control Theory}}.
\newblock \bibinfo{publisher}{Oxford University Press}.

\bibitemdeclare{article}{Boerger-ASMref03}
\bibitem{Boerger-ASMref03}
\bibinfo{author}{E.~B{\"o}rger} (\bibinfo{year}{2003}):
  \emph{\bibinfo{title}{{The ASM Refinement Method}}}.
\newblock {\sl \bibinfo{journal}{{F.A.C.J.}}} \bibinfo{volume}{15}, pp.
  \bibinfo{pages}{237--257}.

\bibitemdeclare{book}{ASM}
\bibitem{ASM}
\bibinfo{author}{E.~B{\"o}rger} \& \bibinfo{author}{R.F. St{\"a}rk}
  (\bibinfo{year}{2003}): \emph{\bibinfo{title}{Abstract State Machines. A
  Method for High Level System Design and Analysis}}.
\newblock \bibinfo{publisher}{Springer}.

\bibitemdeclare{book}{clarke:87}
\bibitem{clarke:87}
\bibinfo{author}{F.~Clarke} (\bibinfo{year}{1987}):
  \emph{\bibinfo{title}{Optimization and Nonsmooth Analysis}}.
\newblock \bibinfo{publisher}{Society for Industrial Mathematics}.

\bibitemdeclare{book}{clarke:97}
\bibitem{clarke:97}
\bibinfo{author}{F.~Clarke}, \bibinfo{author}{Y.~Ledyaev},
  \bibinfo{author}{R.~Stern} \& \bibinfo{author}{P.~Wolenski}
  (\bibinfo{year}{1997}): \emph{\bibinfo{title}{Nonsmooth Analysis and Control
  Theory}}.
\newblock \bibinfo{publisher}{Springer}.

\bibitemdeclare{techreport}{CoSl:08}
\bibitem{CoSl:08}
\bibinfo{author}{J.~Cohen} \& \bibinfo{author}{A.~Slissenko}
  (\bibinfo{year}{2008}): \emph{\bibinfo{title}{Implementation of Timed
  Abstract State Machines with Instantaneous Actions by Machines with Delays}}.
\newblock \bibinfo{type}{Technical Report} \bibinfo{number}{TR-LACL-2008-2},
  \bibinfo{institution}{LACL, University of Paris-12}.

\bibitemdeclare{inproceedings}{COR:95}
\bibitem{COR:95}
\bibinfo{author}{J.~Crow}, \bibinfo{author}{S.~Owre},
  \bibinfo{author}{J.~Rushby}, \bibinfo{author}{N.~Shankar} \&
  \bibinfo{author}{M.~Srivas} (\bibinfo{year}{1995}): \emph{\bibinfo{title}{{A
  Tutorial Introduction to PVS}}}.
\newblock In \bibinfo{editor}{R.~France}, \bibinfo{editor}{S.~Gerhart} \&
  \bibinfo{editor}{M.~Larrondo-Petrie}, editors: {\sl
  \bibinfo{booktitle}{WIFT'95: Workshop on Industrial-Strength Formal
  Specification Techniques}}, \bibinfo{publisher}{IEEE Computer Society Press}.

\bibitemdeclare{book}{DadHou:95}
\bibitem{DadHou:95}
\bibinfo{author}{J.~D'Azzo} \& \bibinfo{author}{C.~Houpis}
  (\bibinfo{year}{1995}): \emph{\bibinfo{title}{Linear Control System Analysis
  and Design: Conventional and Modern}}.
\newblock \bibinfo{publisher}{McGraw Hill}.

\bibitemdeclare{book}{boiten:01}
\bibitem{boiten:01}
\bibinfo{author}{J~Derrick} \& \bibinfo{author}{E~Boiten}
  (\bibinfo{year}{2001}): \emph{\bibinfo{title}{{Refinement in {Z} and
  Object-{Z}: Foundations and Advanced Applications}}}.
\newblock \bibinfo{publisher}{Springer-Verlag UK},
  \doi{10.1007/978-1-4471-0257-1}.

\bibitemdeclare{book}{DorBish:10}
\bibitem{DorBish:10}
\bibinfo{author}{R.~Dorf} \& \bibinfo{author}{R.~Bishop}
  (\bibinfo{year}{2010}): \emph{\bibinfo{title}{Modern Control Systems}}.
\newblock \bibinfo{publisher}{Pearson}.

\bibitemdeclare{inproceedings}{Dut:96}
\bibitem{Dut:96}
\bibinfo{author}{B.~Dutertre} (\bibinfo{year}{1996}):
  \emph{\bibinfo{title}{Elements of Mathematical Analysis in {PVS}}}.
\newblock In: {\sl \bibinfo{booktitle}{TPHOLS 1996}}, {\sl
  \bibinfo{series}{LNCS}} \bibinfo{volume}{1125},
  \bibinfo{publisher}{Springer}.

\bibitemdeclare{book}{DutThBa:97}
\bibitem{DutThBa:97}
\bibinfo{author}{K.~Dutton}, \bibinfo{author}{S.~Thompson} \&
  \bibinfo{author}{B.~Barraclough} (\bibinfo{year}{1997}):
  \emph{\bibinfo{title}{The Art of Control Engineering}}.
\newblock \bibinfo{publisher}{Addison Wesley}.

\bibitemdeclare{book}{FaVi:09}
\bibitem{FaVi:09}
\bibinfo{author}{M.~Fadali} \& \bibinfo{author}{A.~Visioli}
  (\bibinfo{year}{2009}): \emph{\bibinfo{title}{Digital Control Engineering:
  Analysis and Design}}.
\newblock \bibinfo{publisher}{Academic Press}.

\bibitemdeclare{book}{FrPoWo:98}
\bibitem{FrPoWo:98}
\bibinfo{author}{G.~Franklin}, \bibinfo{author}{J.~Powell} \&
  \bibinfo{author}{M.~Workman} (\bibinfo{year}{1996}):
  \emph{\bibinfo{title}{Digital Control Systems}}.
\newblock \bibinfo{publisher}{Prentice Hall}.

\bibitemdeclare{inproceedings}{He:94}
\bibitem{He:94}
\bibinfo{author}{J.~He} (\bibinfo{year}{1994}): \emph{\bibinfo{title}{{From CSP
  to hybrid systems}}}.
\newblock In \bibinfo{editor}{A.W. Roscoe}, editor: {\sl \bibinfo{booktitle}{A
  Classical Mind, Essays in Honour of C.A.R. Hoare}},
  \bibinfo{publisher}{Prentice-Hall International}, pp.
  \bibinfo{pages}{171--189}.

\bibitemdeclare{inproceedings}{Henz:06}
\bibitem{Henz:06}
\bibinfo{author}{T.~A. Henzinger} (\bibinfo{year}{1996}):
  \emph{\bibinfo{title}{{The Theory of Hybrid Automata}}}.
\newblock In: {\sl \bibinfo{booktitle}{Proc. IEEE LICS-96}},
  \bibinfo{publisher}{IEEE}, pp. \bibinfo{pages}{278--292}.
\newblock \bibinfo{note}{{See also
  \url{http://mtc.epfl.ch/~tah/Publications/the\_theory\_of_hybrid_automata.pd%
f}}}.

\bibitemdeclare{misc}{CBTC}
\bibitem{CBTC}
\bibinfo{author}{{IEEE Standard 1474}}: \bibinfo{note}{IEEE Standard for
  Communications-Based Train Control (CBTC) Performance and Functional
  Requirements: IEEE Std 1474.1-2004; IEEE Standard for User Interface
  Requirements in Communications-Based Train Control (CBTC) Systems: IEEE Std
  1474.2-2003; IEEE Recommended Practice for Communications-Based Train Control
  (CBTC) System Design and Functional Allocations: IEEE Std 1474.3-2008}.

\bibitemdeclare{phdthesis}{Jes:05}
\bibitem{Jes:05}
\bibinfo{author}{C.~Jeske} (\bibinfo{year}{2005}):
  \emph{\bibinfo{title}{Algebraic Integration of Retrenchment and Refinement}}.
\newblock Ph.D. thesis, \bibinfo{school}{University of Manchester}.

\bibitemdeclare{book}{Kuo:92}
\bibitem{Kuo:92}
\bibinfo{author}{B.~Kuo} (\bibinfo{year}{1992}): \emph{\bibinfo{title}{Digital
  Control Systems}}.
\newblock \bibinfo{publisher}{Oxford University Press}.

\bibitemdeclare{book}{KAOS}
\bibitem{KAOS}
\bibinfo{author}{A.~van Lamsweerde} (\bibinfo{year}{2009}):
  \emph{\bibinfo{title}{{Requirements Engineering: From System Goals to UML
  Models to Software Specifications}}}.
\newblock \bibinfo{publisher}{Wiley}.

\bibitemdeclare{phdthesis}{Let:01}
\bibitem{Let:01}
\bibinfo{author}{{Letier, E.}} (\bibinfo{year}{2001}):
  \emph{\bibinfo{title}{{Reasoning about Agents in Goal-Oriented Requirements
  Engineering}}}.
\newblock Ph.D. thesis, \bibinfo{school}{{D\'{e}pt. Ing\'{e}nierie
  Informatique, Universit\'{e} Catholique de Louvain}}.

\bibitemdeclare{book}{Ogata:08}
\bibitem{Ogata:08}
\bibinfo{author}{K.~Ogata} (\bibinfo{year}{2008}): \emph{\bibinfo{title}{Modern
  Control Engineering}}.
\newblock \bibinfo{publisher}{Pearson}.

\bibitemdeclare{book}{Para:96}
\bibitem{Para:96}
\bibinfo{author}{P.~Paraskevopoulos} (\bibinfo{year}{1996}):
  \emph{\bibinfo{title}{Digital Control Systems}}.
\newblock \bibinfo{publisher}{Prentice Hall}.

\bibitemdeclare{book}{potter:96}
\bibitem{potter:96}
\bibinfo{author}{B~Potter}, \bibinfo{author}{J~Sinclair} \&
  \bibinfo{author}{D~Till} (\bibinfo{year}{1996}): \emph{\bibinfo{title}{{An
  Introduction to Formal Specification and Z}}}, \bibinfo{edition}{2nd.}
  edition.
\newblock \bibinfo{publisher}{Prentice Hall}.

\bibitemdeclare{misc}{PVS}
\bibitem{PVS}
\bibinfo{author}{{PVS Homepage}}: \bibinfo{note}{\url{http://pvs.csl.sri.com}}.

\bibitemdeclare{article}{RET}
\bibitem{RET}
\bibinfo{author}{{Retrenchment Homepage}}:
  \bibinfo{note}{\url{http://www.cs.man.ac.uk/retrenchment}}.

\bibitemdeclare{book}{roever:98}
\bibitem{roever:98}
\bibinfo{author}{W~P de~Roever} \& \bibinfo{author}{K~Engelhardt}
  (\bibinfo{year}{1998}): \emph{\bibinfo{title}{Data Refinement: Model-Oriented
  Proof Methods and their Comparison}}.
\newblock \bibinfo{publisher}{Cambridge University Press}.

\bibitemdeclare{book}{sekerinski:98}
\bibitem{sekerinski:98}
\bibinfo{author}{E~Sekerinski} \& \bibinfo{author}{K~Sere}
  (\bibinfo{year}{1998}): \emph{\bibinfo{title}{{Program Development by
  Refinement: Case Studies Using the {B}-Method}}}.
\newblock \bibinfo{publisher}{Springer}.

\bibitemdeclare{article}{SlVa:08}
\bibitem{SlVa:08}
\bibinfo{author}{A.~Slissenko} \& \bibinfo{author}{P.~Vasilyev}
  (\bibinfo{year}{2008}): \emph{\bibinfo{title}{{Simulation of Timed Abstract
  State Machines with Predicate Logic model Checking}}}.
\newblock {\sl \bibinfo{journal}{J.U.C.S.}} \bibinfo{volume}{14}, pp.
  \bibinfo{pages}{1984--2006}.

\bibitemdeclare{book}{sontag:98}
\bibitem{sontag:98}
\bibinfo{author}{E.~Sontag} (\bibinfo{year}{1998}):
  \emph{\bibinfo{title}{Mathematical Control Theory}}.
\newblock \bibinfo{publisher}{Springer}.

\bibitemdeclare{inproceedings}{Su:11}
\bibitem{Su:11}
\bibinfo{author}{W.~Su}, \bibinfo{author}{F.~Yang}, \bibinfo{author}{X.~Wu},
  \bibinfo{author}{J.~Gou} \& \bibinfo{author}{H.~Zhu} (\bibinfo{year}{2011}):
  \emph{\bibinfo{title}{{Formal Approaches to Mode Conversion and Positioning
  for Vehicle Systems}}}.
\newblock In: {\sl \bibinfo{booktitle}{{Proc. 3rd IEEE International Workshop
  on Security Aspects of Process and Services Engineering}}}.
\newblock \bibinfo{note}{To appear}.

\bibitemdeclare{book}{Tab:09}
\bibitem{Tab:09}
\bibinfo{author}{P.~Tabuada} (\bibinfo{year}{2009}):
  \emph{\bibinfo{title}{Verification and Control of Hybrid Systems: A Symbolic
  Approach}}.
\newblock \bibinfo{publisher}{Springer}.

\bibitemdeclare{book}{woodcock:96}
\bibitem{woodcock:96}
\bibinfo{author}{J~Woodcock} \& \bibinfo{author}{J~Davies}
  (\bibinfo{year}{1996}): \emph{\bibinfo{title}{{Using Z, Specification,
  Refinement and Proof}}}.
\newblock \bibinfo{publisher}{Prentice Hall}.

\end{thebibliography}

\end{document}